\documentclass[prc,superscriptaddress,unsortedaddress,twocolumn,showpacs]{revtex4}
\usepackage{graphicx}
\usepackage{amsmath}
\usepackage{amssymb}
\usepackage{times}
\usepackage{bm}
\usepackage{color}
\usepackage{ulem}

\def\fun#1#2{\lower3.6pt\vbox{\baselineskip0pt\lineskip.9pt
\ialign{$\mathsurround=0pt#1\hfil##\hfil$\crcr#2\crcr\sim\crcr}}}
\newcommand{\beq}{\begin{equation}}
\newcommand{\eeq}{\end{equation}}
\newcommand{\bea}{\begin{eqnarray}}
\newcommand{\eea}{\end{eqnarray}}

\begin{document}

\title{Applicability of the continuum-discretized coupled-channels
method to the deuteron breakup at low energies}

\author{Kazuyuki Ogata}
\email[]{kazuyuki@rcnp.osaka-u.ac.jp}
\affiliation{Research Center for Nuclear Physics (RCNP), Osaka
University, Ibaraki 567-0047, Japan}

\author{Kazuki Yoshida}
\affiliation{Research Center for Nuclear Physics (RCNP), Osaka
University, Ibaraki 567-0047, Japan}

\date{\today}

\begin{abstract}
We re-examine the deuteron elastic breakup cross sections on $^{12}$C
and $^{10}$Be at low incident energies, for which a serious discrepancy between
the continuum-discretized coupled-channels method (CDCC)
and the Faddeev--Alt-Grassberger-Sandhas theory (FAGS) was
pointed out.
We show the closed-channels neglected in the preceding study
affect significantly the breakup cross section calculated with CDCC,
resulting in good agreement with the result of FAGS.
\end{abstract}


\maketitle

{\it Introduction.}
Projectile breakup reactions have played a major role in studying
the structure of loosely-bound nuclei~\cite{Tan13}.
Such a reaction contains at least three particles in the final state.
Thus, one may say that the accurate description of the three-body
breakup process is a minimum requirement for nuclear reaction theories.
It is well known that the Faddeev theory~\cite{Fad60},
or, alternatively, the Alt-Grassberger-Sandhas (AGS) theory~\cite{Alt67}
gives the exact solution to such a three-body scattering problem.
On the other hand, the continuum-discretized coupled-channels
method (CDCC)~\cite{Kam86,Aus87,Yah12} has widely been applied
with high success to projectile breakup reactions at various incident
energies. The theoretical foundation of CDCC was given in
Refs.~\cite{Aus89,Aus96} in connection with the distorted-wave Faddeev
formalism~\cite{BR82}.
Quite recently~\cite{Del07}, invention of the treatment of the Coulomb interaction
made the Faddeev-AGS theory (FAGS) applicable to various three-body
breakup reactions, and the results of FAGS have directly been compared
with those of CDCC. In many cases the two give very similar cross sections,
which validates CDCC as an effective three-body reaction model, as predicted
in Refs.~\cite{Aus89,Aus96}.

In a systematic comparison~\cite{Upa12} between FAGS and CDCC,
however, it was shown that at high incident energies $E_d$ of deuteron,
($d,p$) transfer cross sections calculated with CDCC somewhat deviate from
those with FAGS, i.e., the exact cross sections.
More seriously, at $E_d$ below about 20~MeV, the deuteron
elastic breakup cross sections obtained with CDCC overshoot those of
FAGS by about a factor of three at most.
The latter finding can particularly be a striking indication of the
limitation of CDCC, suggesting that at low incident energies
one has to rely on a more elaborated
reaction model or exact FAGS for describing even elastic breakup processes.
In Ref.~\cite{Upa12}, however, the so-called closed channels (see below)
were not included. As mentioned in literature, e.g., Refs.~\cite{Aus87,Aus96},
inclusion of closed channels is crucial for quantitative discussion
on observables, at low incident energies in particular.
This has numerically been confirmed in Ref.~\cite{AV10} for a one-dimensional
scattering problem, and in Ref.~\cite{DD12} for the scattering of $^{11}$Be.
There exist several indications of the importance of closed channels
also for transfer reactions~\cite{Oga03,Fuk11,Fuk15}.
Under the circumstances, in the present study, we revisit the problem reported on the
low-energy elastic breakup cross sections for $^{10}$Be($d,pn$)$^{10}$Be at
$E_d=21$~MeV and $^{12}$C($d,pn$)$^{12}$C at $E_d=12$~MeV,
and discuss more in detail the convergence of CDCC results,
putting emphasis on the closed channels.

{\it CDCC and closed-channels.}
We give a brief review on CDCC; for more details, see,
e.g., Refs.~\cite{Kam86,Aus87,Yah12}.
We describe the deuteron elastic breakup with the target
nucleus $A$, on the basis of a $p+n+A$ three-body model.
We do not explicitly take into account
the excitation of $A$ during the breakup process.
We neglect also the intrinsic spin of each
of the three particles, following Ref.~\cite{Upa12}.
In CDCC the total three-body wave
function for the total angular momentum $J$ and its projection $M$
is expanded in term of
the complete set of the projectile wave function $\{ \phi \}$:
\bea
\Psi^{JM}({\bm r},{\bm R})
&=&
\sum_{i=0}^{i_{\rm max}}
\sum_{\ell=0}^{\ell_{\rm max}}
\sum_{L=|J-\ell|}^{J+\ell}
\phi_{i\ell}(r) \chi_c(R)
\nonumber \\
& &
\times
\left[
i^\ell Y_\ell (\hat{\bm r})\otimes i^L Y_L (\hat{\bm R})
\right]_{JM}
,
\label{cdcc}
\eea
where ${\bm r}$ (${\bm R}$) is the coordinate of $p$ (the center-of-mass
of $d$) relative to $n$ ($A$).
$i$ is the energy index and $i=0$ represents the ground state of $d$.
The orbital angular momenta corresponding to ${\bm r}$ and ${\bm R}$
are denoted by $\ell$ and $L$, respectively; $Y_{lm}$ is the spherical harmonics.
We have put the channel indices of the scattering wave $\chi$ altogether
in $c$, i.e., $c=\{J,i,\ell,L\}$.
In the derivation of Eq.~(\ref{cdcc}) we have discretized the $p$-$n$ continua
with the so-called momentum-bin average method:
\beq
\phi_{i\ell}(r)
=\dfrac{1}{\sqrt{\Delta k}}
\int_{k_i}^{k_i+\Delta k} dk \;
\varphi_{k,\ell}(r),
\eeq
where $k_i= (i-1) \Delta k$ and $\varphi_{k}$ is the partial wave of the $p$-$n$ scattering wave function under a $p$-$n$ interaction $V_{pn}$, with $k$ the absolute value of the asymptotic relative
momentum.
The discretized $p$-$n$ energy of the $i$th state ($i>0$)
is given by~\cite{Kam86}
\[
\hat{\epsilon}_i = \dfrac{\hbar^2}{2 \mu_{pn}}
\left[
\dfrac{\Delta k}{12}
+\dfrac{(2k_i + \Delta k)^2}{4}
\right],
\]
where $\mu_{pn}$ is the $p$-$n$ reduced mass.
The size $\Delta k$ of the momentum bin, the maximum linear momentum
$k_{\rm max}=i_{\rm max} \Delta k$ (in the unit of $\hbar$),
and $\ell_{\rm max}$
are key values for determining the reaction model-space of CDCC.

\begin{figure}[b]
\begin{center}
 \includegraphics[width=0.45\textwidth]{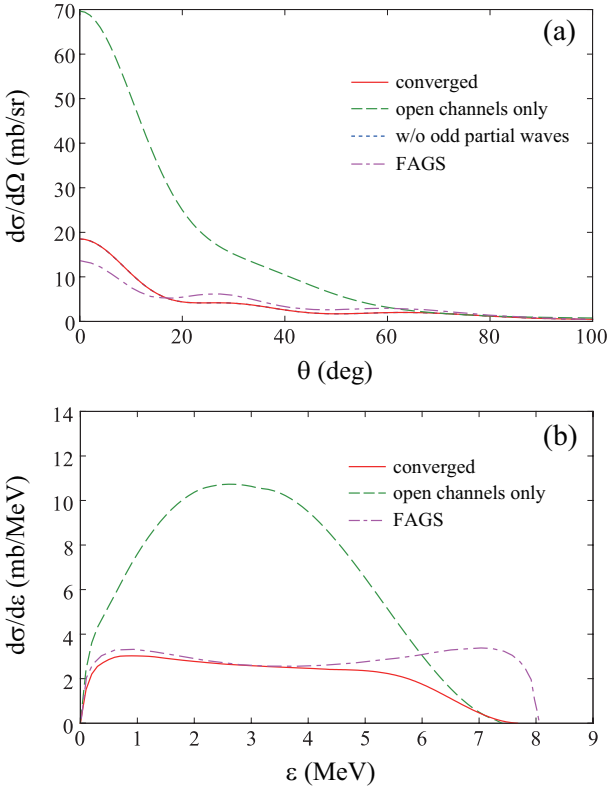}
 \caption{
  (Color online) (a) Angular distribution and (b) breakup energy distribution
  of the elastic breakup cross  section for $^{12}$C($d,pn$)$^{12}$C at $E_d=12$~MeV.
  The solid, dashed, and dash-dotted lines in each panel show the converged CDCC result,
  the result of CDCC calculated with including only the open channels, and
  the result of FAGS taken from Ref.~\cite{Upa12}, respectively.
  The dotted line in (a) is the same as the solid line but with omitting the odd
  partial waves between $p$ and $n$.
  }
 \label{fig1}
\end{center}
\end{figure}
The asymptotic form of $\chi_c$ is given by
\beq
\chi_c
\to
U^{(-)}_{L,\eta_{i}} (K_{i} R)\delta_{cc_0}
-\sqrt{K_{0}/K_{i}}\;
S_{cc_0}
U^{(+)}_{L,\eta_{i}} (K_{i} R)
\label{bc1}
\eeq
for $E_{i} > 0$, and
\beq
\chi_c
\to
-S_{cc_0}
W_{-\eta_{i},L+1/2} (-2i K_{i} R)
\label{bc2}
\eeq
for $E_{i} \le 0$, where $E_{i}=E-\hat{\epsilon}_i$ and
$K_{i}=\sqrt{2 \mu E_i}/\hbar$; $c_0$ represents the incident channel.
$U^{(-)}_{L,\eta_{i}}$ ($U^{(+)}_{L,\eta_{i}}$) is the
incoming (outgoing) Coulomb wave function with the Sommerfeld
parameter $\eta_{i}$ and
$W_{-\eta_{i},L+1/2}$ is the Whittaker function.
Channels having  $E_{i} > 0$ and $E_{i} \le 0$ are called
open channels and closed channels, respectively.
$S_{cc_0}$ for open channels are scattering matrix elements,
with which physics observables are calculated in a standard manner.
On the other hand, $S_{cc_0}$ for closed channels
are not related to observables,
at least directly. It is obvious, however, that the closed channels
can affect the breakup observables through mainly continuum-continuum
couplings~\cite{Aus96}.

\begin{figure}[b]
\begin{center}
 \includegraphics[width=0.45\textwidth]{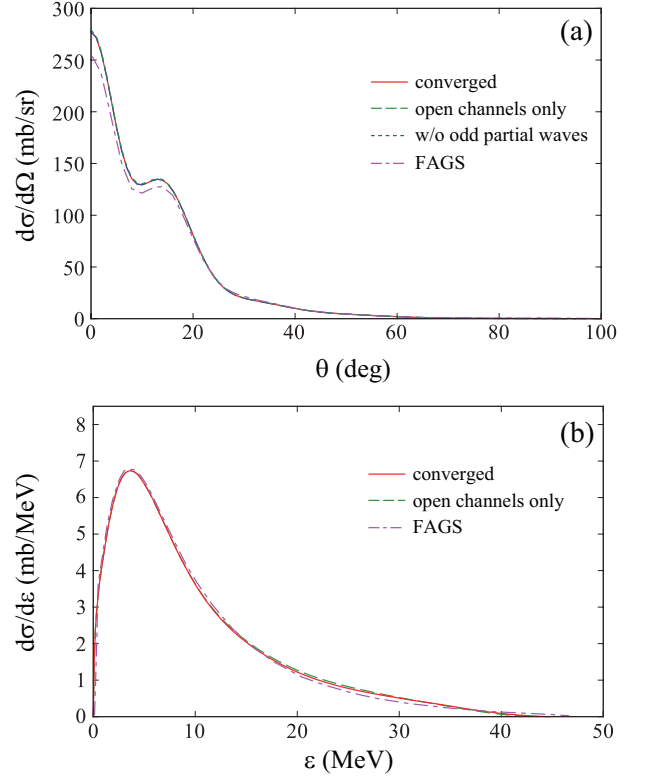}
 \caption{
  (Color online) Same as Fig.~\ref{fig1} but at $E_d=56$~MeV.
  }
 \label{fig2}
\end{center}
\end{figure}

\begin{figure}[t]
\begin{center}
 \includegraphics[width=0.45\textwidth]{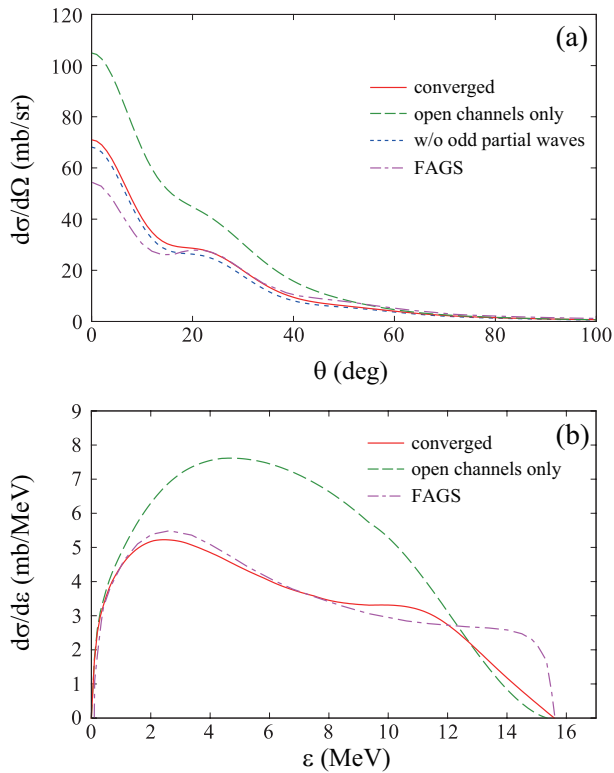}
 \caption{
  (Color online) Same as Fig.~\ref{fig1} but for
  $^{10}$Be($d,pn$)$^{10}$Be at $E_d=21$~MeV.
  }
 \label{fig3}
\end{center}
\end{figure}

{\it Results and discussion.}
In the CDCC calculation shown below, we disregard the intrinsic
spins of $p$ and $n$ as mentioned, and also the Coulomb breakup.
For $V_{pn}$, we adopt the one-range Gaussian interaction of
Ref.~\cite{Ohm70}, and for the nucleon-nucleus optical potential,
we employ the CH89 global potential~\cite{Var91}.
These are the same model setting as in Ref.~\cite{Upa12}.
We use $\Delta k=0.05$~fm$^{-1}$ and $\ell_{\rm max}=8$ for all the
calculation shown below. As for $k_{\rm max}$, we take
$0.9$~fm$^{-1}$ for $^{12}$C($d,pn$)$^{12}$C at $E_d=12$~MeV
(Fig.~\ref{fig1}) and 1.1~fm$^{-1}$ for other two reactions
(Figs.~\ref{fig2} and \ref{fig3}).
We have checked the convergence of the breakup cross sections
by further increasing the model space, and thereby convergence with
98\% accuracy has been confirmed.
In the multipole expansion of the nucleon-nucleus optical potential,
we take the multipolarities $\lambda$ up to 16; it turned out
that the multipoles for $\lambda > 8$ have no effect on the results
shown below.

Figure~\ref{fig1}(a) shows the angular distribution of the
deuteron breakup cross section on $^{12}$C at $E_d=12$~MeV
integrated over the $p$-$n$ breakup energy $\epsilon$. The horizontal
axis is the scattering angle $\theta$ of the center-of-mass of the  $p$-$n$ system.
The solid line is the converged result of CDCC that agrees
well with the result of FAGS (dash-dotted line)
taken from Fig.~9(a) of  Ref.~\cite{Upa12}.
The dashed line in Fig.~\ref{fig1}(a) is the CDCC result calculated
with including open channels only, as in Ref.~\cite{Upa12},
which seems to be inside the hatched band in Fig.~9(a) of Ref.~\cite{Upa12}.
One sees in Fig.~\ref{fig1}(a) a significant reduction of the cross section
due to the coupling with the closed channels. Although still
a small difference remains between the converged CDCC
in the present study and the FAGS results in Ref.~\cite{Upa12},
we conclude that the severe overshooting problem of CDCC pointed out in
Ref.~\cite{Upa12} is mainly due to the lack of the closed channels
in the CDCC calculation.
The dotted line in Fig.~\ref{fig1}(a) shows the converged result
of CDCC including only the even partial waves of $\ell$, which
perfectly agrees with the solid line. This is due to the neglect
of the Coulomb breakup and to the small difference between the
$p$-$^{12}$C and $n$-$^{12}$C potentials. This fact allows one to
neglect the odd partial waves in CDCC, at least in some cases, which
will make the comparison between CDCC and FAGS much easier, although in
reality we always have the Coulomb breakup effect.
Figure~\ref{fig1}(b) is the $p$-$n$ breakup energy distribution,
with $\theta$ integrated. The features
of the results are the same as in Fig.~\ref{fig1}(a).
The disagreement found in the high $\epsilon$ region will need
further investigations.

Next we show in Fig.~\ref{fig2} the results for $^{12}$C at $E_d=56$~MeV.
For this reaction, no significant difference between CDCC and FAGS was
reported in Ref.~\cite{Upa12}. It is quite natural that the coupling to the closed-channels is
less important at higher incident energy. One can clearly see this for
both angular distribution (Fig.~\ref{fig2}(a)) and breakup energy
distribution (Fig.~\ref{fig2}(b)). In fact, the adopted $k_{\rm max}$
($0.9$~fm$^{-1}$) for this reaction that gives convergence is very
close to the threshold of the open channels, $1.05$~fm$^{-1}$.
It is thus quite trivial that the two lines agree with each other in
Fig.~\ref{fig2}(a) and Fig.~\ref{fig2}(b). In any case, checking
the convergence with respect to $k_{\rm max}$ is necessary.

Figure~\ref{fig3} is the result for
$^{10}$Be($d,pn$)$^{10}$Be at $E_d=21$~MeV.
The role of the closed channels and the
agreement between the converged CDCC and FAGS are
the same as in Fig.~\ref{fig1}, although
the role of the odd partial waves is appreciable in this
reaction.

{\it Summary.}
We have reinvestigated deuteron elastic breakup reactions on
$^{12}$C and $^{10}$Be at low incident energies, in which
significant difference in the cross sections between CDCC
and FAGS was reported~\cite{Upa12}. We checked carefully the convergence
of CDCC, with respect to the maximum $p$-$n$ breakup momentum $k_{\rm max}$
in particular. The crucial importance of the closed channels
was shown, and the converged CDCC results agree well
with the FAGS results shown in Ref.~\cite{Upa12}. At higher energy, the
closed channels turned out to less important, as expected.

In conclusion, we have demonstrated the applicability of CDCC
to elastic breakup reactions $^{10}$Be($d,pn$)$^{10}$Be at
$E_d=21$~MeV and $^{12}$C($d,pn$)$^{12}$C at $E_d=12$~MeV,
by confirming the convergence of the CDCC model space with respect to $k_{\rm max}$.
As a next step, a more systematic investigation on the role of closed channels,
in transfer reactions in particular, will be important.

The authors thank A. M. Moro, P. Capel, and F. M. Nunes for their valuable
comments on the manuscript.
This work was supported in part by Grants-in-Aid of the Japan Society
for the Promotion of Science (Grants No. JPT16K053520 and No. JP15J01392)
and by the ImPACT Program of the Council for Science, Technology and
Innovation (Cabinet Office, Government of Japan).
The computation was carried out with the computer facilities
at the Research Center for Nuclear Physics, Osaka University.



\begin{thebibliography}{00}

\bibitem{Tan13}
I. Tanihata, H. Savajols, and R. Kanungo, Prog. Part. Nucl. Phys. {\bf 68},
215 (2013) and
references cited therein.

\bibitem{Fad60}
L.~D.~Faddeev, Zh. Eksp. Theor. Fiz. {\bf 39}, 1459 (1960)
[Sov. Phys. JETP {\bf 12}, 1014 (1961)].

\bibitem{Alt67}
E.~O.~Alt, P.~Grassberger, and W.~Sandhas, Nucl. Phys. B {\bf 2}, 167 (1967).

\bibitem{Kam86}
M.~Kamimura, M.~Yahiro, Y.~Iseri, Y.~Sakuragi, H.~Kameyama, and M.~Kawai,
Prog. Theor. Phys. Suppl. {\bf 89}, 1 (1986).

\bibitem{Aus87}
N.~Austern, Y.~Iseri, M.~Kamimura, M.~Kawai, G.~Rawitscher, and M.~Yahiro,
Phys. Rep. {\bf 154}, 125 (1987).

\bibitem{Yah12}
M.~Yahiro, K.~Ogata, T.~Matsumoto, and K.~Minomo,
Prog. Theor. Exp. Phys. {\bf 2012}, 01A206 (2012).

\bibitem{Aus89}
N.~Austern, M.~Yahiro, and M.~Kawai,
Phys.\ Rev.\ Lett. {\bf 63}, 2649 (1989).

\bibitem{Aus96}
N.~Austern, M.~Kawai, and  M.~Yahiro,
Phys. Rev. C {\bf 53}, 314 (1996).

\bibitem{BR82}
M.C. Birse and E. F. Redish, Nucl. Phys. A{\bf 406}, 149 (1982).

\bibitem{Del07}
A. Deltuva, A. M.~Moro, E.~Cravo, F. M.~Nunes, and A. C.~Fonseca,
Phys.\ Rev.\ C {\bf 76}, 064602 (2007).

\bibitem{Upa12}
N. J. Upadhyay, A. Deltuva, and F. M. Nunes,
Phys.\ Rev.\ C {\bf 85}, 054621 (2012).

\bibitem{AV10}
N. Ahsan and A. Volya,
Phys.\ Rev.\ C {\bf 82}, 064607 (2010).

\bibitem{DD12}
T. Druet and P. Descouvemont,
Eur. Phys. J. A {\bf 48}, 147 (2012).

\bibitem{Oga03}
K. Ogata, M. Yahiro, Y. Iseri, and M. Kamimura,
Phys.\ Rev.\ C {\bf 67}, 011602(R) (2003).

\bibitem{Fuk11}
T. Fukui, K. Ogata, and M. Yahiro,
Prog. Theor. Phys. {\bf 125}, 1193 (2011).

\bibitem{Fuk15}
T. Fukui, K. Ogata, and M. Yahiro,
Phys.\ Rev.\ C {\bf 91}, 014604 (2015).

\bibitem{Ohm70}
T. Ohmura, B. Imanishi, M. Ichimura, and M. Kawai,
Prog. Theor. Phys. \textbf{43}, 347 (1970).

\bibitem{Var91}
R. L. Varner {\it et al.}, Phys. Rep. {\bf 201}, 57 (1991).

\end{thebibliography}
\end{document}